\def\b{\bibitem}
\def\bpi{\bbox{\pi}}
\documentstyle[aps,prl,psfig,floats]{revtex} 
\draft
\begin{document}
\def\SNG{{\em Physical Review Style and Notation Guide}}
\def\LUG {{\em \LaTeX{} User's Guide \& Reference Manual}}
\def\btt#1{{\tt$\backslash$\string#1}}%
\def\REVTeX{REV\TeX}
\def\AmS{{\protect\the\textfont2
        A\kern-.1667em\lower.5ex\hbox{M}\kern-.125emS}}
\def\AmSLaTeX{\AmS-\LaTeX}
\def\BibTeX{\rm B{\sc ib}\TeX}
\twocolumn[\hsize\textwidth\columnwidth\hsize\csname@twocolumnfalse%
\endcsname
\title{Nonanalytic Magnetization Dependence of the Magnon Effective Mass \\ 
          in Itinerant Quantum Ferromagnets}
\author{D. Belitz}
\address{Department of Physics and Materials Science Institute,
University of Oregon,
Eugene, OR 97403}
\author{T.~R. Kirkpatrick}
\address{Institute for Physical Science and Technology, and Department of Physics\\
University of Maryland, College Park, MD 20742}
\author{A.~J. Millis}
\address{Department of Physics and Astronomy, The Johns Hopkins University\\
         3400 North Charles Street, Baltimore, MD 21218}
\author{Thomas Vojta}
\address{Department of Physics and Materials Science Institute,
         University of Oregon, Eugene, OR 97403\\
         and
         Institut f{\"u}r Physik, TU Chemnitz-Zwickau, D-09107 Chemnitz, FRG}
\date{\today}
\maketitle
\begin{abstract}
The spin wave dispersion relation in both clean and disordered itinerant
quantum ferromagnets is calculated. It is found that effects akin to
weak-localization physics cause the frequency of the spin-waves to be a 
{\em nonanalytic} function of the magnetization $m$. For low frequencies 
$\Omega$, small wavevectors ${\mathbf k}$, and $m\rightarrow 0$, the 
dispersion relation is found to be of the form 
$\Omega = {\rm const.}\times m^{1-\alpha}\,{\mathbf k}^2$, 
with $\alpha = (4-d)/2$ ($2<d<4$) for disordered systems, and 
$\alpha = (3-d)$ ($1<d<3$) for clean ones. In $d=4$ (disordered) and 
$d=3$ (clean), $\Omega\propto m\ln(1/m)\,{\mathbf k}^2$. 
Experiments to test these predictions are proposed.
%
%
\end{abstract}
]
One of the best known examples of quantum long-range order is the
ferromagnetic (FM) state in itinerant electron systems at zero 
temperature ($T=0$). An important manifestation of this order is 
the existence of spin waves\cite{Forster}.
In conventional Heisenberg ferromagnets the damping of the spin wave is
negligible, and the dispersion relation has the
form\cite{Moriya},
\begin{equation}
\Omega = D(m)\,{\mathbf k}^{2} + o(\vert{\mathbf k}\vert^2)\quad,
\label{eq:1}
\end{equation}
with $o(\epsilon)$ denoting terms that are smaller than $\epsilon$.
The coefficient $D$ depends on the dimensionless magnetization 
$m = (n_{\uparrow}-n_{\downarrow})/n$, with $n_{\uparrow}$ and
$n_{\downarrow}$ the densities of spin-up and spin-down electrons,
respectively, and $n = n_{\uparrow} + n_{\downarrow}$. In the conventional
theory for clean `weak ferromagnets'\cite{Moriya}, 
$D(m\rightarrow 0) = D_0\,m$.
$D_0\approx v_{\rm F}/k_{\rm F}$, with $k_{\rm F}$ the Fermi wavenumber and
$v_{\rm F} = k_{\rm F}/\mu$ the Fermi velocity, is on the order of the 
inverse of the electron mass $\mu$, and Eq.\ (\ref{eq:1}) is valid for 
$\vert{\mathbf k}\vert < k_{\rm F}\,m\ll k_{\rm F}$. We
will show below that these results do {\em not}
correctly describe the small-$m$ behavior of metallic ferromagnets.

A crucial assumption in the derivation of Eq.\ (\ref{eq:1}) is that the
interactions between spin fluctuations are short-ranged. This assumption
is of doubtful validity in the context of itinerant ferromagnets, since
in metals at $T=0$ there exist soft modes that can couple to the spin
fluctuations and lead to an effective long-ranged interaction.
Indeed, recent work on the $T=0$ FM phase
transition in both disordered\cite{us2} and clean\cite{us3} itinerant
electron systems has shown that in spatial
dimensions $d=2,3$ the asymptotic critical behavior is largely
determined by the coupling of non-critical soft modes to the critical
spin fluctuations. In disordered systems, these soft modes are the same 
`diffusons' that cause the so-called
weak-localization effects\cite{WeakLocalizationFootnote}. In clean
systems, they are the usual particle-hole excitations that lead to the
well-known nonanalyticities in Fermi liquids\cite{BaymPethick} which have
recently been shown to be the clean analogues of the weak-localization 
effects\cite{fermions}. These non-critical soft modes
cause the critical spin fluctuations to interact via dimensionality
dependent long-range effective forces.
In the paramagnetic phase, the same
physics is known to lead to a nonanalyticity in the wavenumber dependent
spin susceptibility of the form 
\begin{equation}
\chi_s({\mathbf k}) \sim {\rm const.} + \vert{\mathbf k}\vert^{\zeta}\quad, 
\label{eq:2}
\end{equation}
with $\zeta = d-2$ (disordered) and $\zeta = d-1$ (clean), 
respectively\cite{chi_s}.

In this Letter we consider the FM phase, and show that the long-ranged spin
interactions that are mediated by the diffusons, or their clean counterparts,
render invalid the standard results for
the magnon dispersion. We find that a non-zero magnetization cuts off
the long-ranged interaction at a scale $\ell_m \sim m^{-1}$ (clean) or
$\ell_m \sim m^{-1/2}$ (disordered), which transforms the singular
dependence on the wavenumber into one on the magnetization. The
magnon dispersion is then given by Eq.\ (\ref{eq:1}), but with a
nonanalytic $m$-dependence of $D$.
For disordered electronic systems, we find
\begin{mathletters}
\label{eqs:3}
\begin{eqnarray}
D(m\rightarrow 0)&=& c_d\,m\,\left[m^{-(4-d)/2} + O(1)\right]
                                         \ ,\  (2<d<4)\quad,
\nonumber\\ 
D(m\rightarrow 0)&=& c_4\,m\,\left[\ln (1/m) + O(1)\right]
                                         \ ,\  (d=4)\quad,
\label{eq:3a}
\end{eqnarray}
and $D(m\rightarrow 0) \sim m$ for $d>4$. For clean systems,
\begin{eqnarray}
D(m\rightarrow 0)&=& {\tilde c}_d\,m\,\left[m^{-(3-d)} + O(1)\right]
                                               \ ,\  (1<d<3)\quad, 
\nonumber\\
D(m\rightarrow 0)&=& {\tilde c}_3\,m\,\left[\ln (1/m) + O(1)\right]
                                                  \ ,\  (d=3)\quad,  
\label{eq:3b}
\end{eqnarray}
\end{mathletters}%
and $D(m\rightarrow 0) \sim m$ for $d>3$. In these equations, $c_d$ and 
${\tilde c}_d$ are positive constants.

In the remainder of this Letter we derive and further discuss these results.
For simplicity, we consider a $d$-dimensional continuum model of interacting
clean or disordered electrons\cite{ModelFootnote}, 
and pay particular attention to the particle-hole
spin-triplet contribution to the electron-electron interaction term in the
action, whose (repulsive) coupling constant we denote by $\Gamma_t$.
Writing only the latter interaction term explicitly, and denoting the 
spin density by ${\bf n}_s,$ the action reads
\begin{equation}
S = S_{0} + S_{\rm int}^{\,t} = S_{0} + \frac{\Gamma_t}{2}\int dx\,
     {\mathbf n}_s(x)\cdot{\mathbf n}_{s}(x)\quad,  
\label{eq:4}
\end{equation}
where $S_0$ contains all contributions to the action other than
$S_{\rm int}^{\,t}$. In particular, it contains the 
particle-hole spin-singlet and
particle-particle interactions, as well as the coupling to the disorder. 
$\int dx\equiv\int d{\mathbf x}\int_0^{1/T}d\tau$, and we use a $(d+1)$-vector
notation $x\equiv({\mathbf x},\tau)$, with 
${\mathbf x}$ a vector in real space,
and $\tau$ the imaginary time. We perform a Hubbard-Stratonovich decoupling
of $S_{\rm int}^{\,t}$ by introducing a classical vector field ${\mathbf M}(x)$ 
with components $M_i$ ($i=1,2,3$) that couples to ${\mathbf n}_s(x)$ and whose
average is proportional to the magnetization, and we integrate out all
fermionic degrees of freedom\cite{Hertz}. In this way we obtain the
partition function in the form
\begin{mathletters}
\label{eqs:5}
\begin{equation}
Z=e^{-F_0/T}\int D[{\mathbf M}]\,\exp [-\Phi ({\mathbf M})]\quad.
\label{eq:5a}
\end{equation}
Here $F_{0}$ is the part of the free energy that does not depend on the
magnetization, and $\Phi$ is a Landau-Ginzburg-Wilson (LGW) functional,
\begin{eqnarray}
\Phi ({\mathbf M})&=&\frac{\Gamma_t}{2}\int dx\,{\mathbf M}(x)\cdot{\mathbf M}
                                                                     (x)
\nonumber\\
&&-\ln\left\langle\exp\left[-\Gamma_t\int dx\,{\mathbf M}(x)\cdot
            {\mathbf n}_s(x)\right]\right\rangle_{S_0}\quad,
\label{eq:5b}
\end{eqnarray}
\end{mathletters}%
where $\left\langle\ldots\right\rangle_{S_0}$ denotes an average
taken with respect to the reference action $S_0$.

Next, we expand in fluctuations about the ordered state. In order to ensure
that the $O(3)$ symmetry is still manifest in the ordered state,
we write\cite{ZJ}
\begin{mathletters}
\label{eqs:6}
\begin{equation}
{\mathbf M}(x) = \rho (x)\,\bbox{\hat\phi}(x)\quad, 
\label{eq:6a}
\end{equation}
with $\rho (x)$ the amplitude of ${\mathbf M}(x)$, and $\bbox{\hat\phi}(x)$ 
a unit vector,
\begin{equation}
\bbox{\hat\phi}^{2}(x) = 1\quad. 
\label{eq:6b}
\end{equation}
\end{mathletters}%
Further, we take the system to be ordered in the $3$-direction and
parametrize $\bbox{\hat\phi}$ and $\rho$ by
\begin{mathletters}
\label{eqs:7}
\begin{equation}
\bbox{\hat\phi} = (\bpi,\sigma)\quad, 
\label{eq:7a}
\end{equation}
with $\bpi = (\pi_1,\pi_2)$, $\sigma^2 = 1 - \bpi^2$, and
\begin{equation}
\rho (x) = m + \delta\rho (x)\quad,
\label{eq:7b}
\end{equation}
\end{mathletters}%
with $m=\left\langle\rho (x)\right\rangle$ proportional to the magnetization.
According to Goldstone's theorem, the transverse fluctuations
$\bpi(x)$ are soft, or of long range. 
$\Phi({\mathbf M})$ can then be expanded in the fluctuations $\delta\rho$ 
and $\bpi$ as,
\begin{mathletters}
\label{eqs:8}
\begin{equation}
\Phi ({\mathbf M}) = \Phi (m\bbox{\phi}_3) + \delta\Phi ({\mathbf M})\quad,
\label{eq:8a}
\end{equation}
with $\bbox{\phi}_3$ a unit vector in $3$-direction, and 
\begin{eqnarray}
\delta\Phi ({\mathbf M}) &=& \frac{\Gamma_t}{2}\int dx\,\left[\rho^2(x)
                                                               - m^2\right] 
\nonumber\\
&&-\log \left\langle e^{-\Gamma_t\int dx\,
   {\mathbf M}(x)\cdot{\mathbf n}_s(x) - m\,n_{s,3}(x)}\right\rangle_{S'_0}
                                                             \ ,
\label{eq:8b}
\end{eqnarray}
with
\begin{equation}
S'_0 = S_0 - \Gamma_t\,m\int dx\,n_{s,3}(x)\quad. 
\label{eq:8c}
\end{equation}
\end{mathletters}%

The correlation functions in Eq.\ (\ref{eq:8b}) that one obtains by
expanding the exponential determine the coefficients in the LGW functional.
They are correlation functions of a reference ensemble whose action is 
given by Eq.\ (\ref{eq:8c}), which
describes the reference ensemble $S_0$ in an external magnetic
field given by $-\Gamma_t m$. Here we are interested in the transverse spin
susceptibility, which can be obtained from the imaginary frequency
correlation function,
\begin{equation}
\chi_t({\mathbf k},\Omega_n) = 
      \left\langle\vert\pi_1({\mathbf k},\Omega_n)\vert^2
                             \right\rangle\quad ,  
\label{eq:9}
\end{equation}
with $\Omega_n$ a bosonic Matsubara frequency. Let us first
consider the terms in Eq.\ (\ref{eq:8b}) that are bilinear in
$\bpi$, which we denote by $\delta\Phi_{\pi\pi}$.
We further integrate out $\rho(x)$ in saddle-point approximation, i.e.
we neglect the fluctuations $\delta\rho$. We will justify this procedure
later, and also discuss terms of higher order in $\bpi$.
Taylor expanding Eq.\ (\ref{eq:8b}) gives,
\begin{mathletters}
\label{eqs:10}
\begin{equation}
\delta\Phi_{\mathbf\pi\pi} = \frac{\Gamma_tm}{2}\int dxdy
          \sum_{i,j=1}^{2}\pi_{i}(x)\,K_{ij}(x,y)\,\pi_{j}(y)\quad, 
\label{eq:10a}
\end{equation}
with,
\begin{eqnarray}
K_{ij}(x,y)&=&-\Gamma_tm\left\langle n_{s,i}(x)\,n_{s,j}
           (y)\right\rangle_{S'_0}
\nonumber\\
&&-\delta (x-y)\,\delta_{ij}\langle n_{s,3}(x)\rangle_{S'_0}\quad.  
\label{eq:10b}
\end{eqnarray}
In this approximation,
\begin{equation}
\chi_t({\mathbf k},\Omega_n) = \frac{K_{11}({\mathbf k},\Omega_n)}{\Gamma_tm}\,
    \left[ K_{11}^2({\mathbf k},\Omega_n) 
         - K_{12}^2({\mathbf k},\Omega_n)\right]^{-1}\ , 
\label{eq:10c}
\end{equation}
i.e., the kernel $K_{ij}$ determines the spin wave
spectrum. Note the Goldstone mode structure of this result:
Taking the Fourier transform of Eq.\ (\ref{eq:10b}), we have ($i=1,2$)
\begin{equation}
K_{ij}({\mathbf k},\Omega_n) = 
    -\Gamma_tm\,\chi_{ij}^{\rm (ref)}({\mathbf k},\Omega_n) 
         - \delta_{ij}\left\langle n_{s,3}\right\rangle_{S'_0}\quad.
\label{eq:10d}
\end{equation}
\begin{equation}
\chi_{ij}^{\rm (ref)}({\mathbf k},\Omega_n) = 
    \left\langle n_{s,i}({\mathbf k},\Omega_n)\,
         n_{s,j}(-{\mathbf k},-\Omega_n)\right\rangle_{S'_0}\quad,
\label{eq:10e}
\end{equation}
\end{mathletters}%
is the transverse part of the spin susceptibility in the reference
ensemble with action $S'_0$. A Ward identity that relates
the reference system's magnetization to its static, homogeneous transverse
spin susceptibility\cite{ZJ} ensures that $K_{ij}(0,0) = 0$, i.e., 
transverse excitations are soft.

Expanding in powers of the frequency, one finds
\begin{equation}
\chi_{ij}^{\rm (ref)}({\mathbf k},\Omega_n) = 
  \delta_{ij}\chi_t^{\rm (ref)}({\mathbf k})
    - \frac{ic}{m}\,\vert\Omega_n\vert\,
             \left[\delta_{i 1}\delta_{j 2} 
          + \delta_{i 2}\delta_{j 1}\right],
\label{eq:11}
\end{equation}
with $c \propto \mu^2/k_{\rm F}$ a constant. 
In the absence of weak-localization effects, one would
have $\chi_t^{\rm (ref)}({\mathbf k}) = \chi_t - {\tilde c}{\mathbf k}^2$, 
with ${\tilde c}$ another constant independent of $m$. However, 
due to weak-localization effects in disordered systems, and their analogues
in clean ones, $\chi_t^{\rm (ref)}$ has a singularity
at ${\mathbf k} = m = 0$. For $m\equiv 0$ this has been shown using
perturbation theory\cite{chi_s} as well as more general 
renormalization group (RG) 
arguments\cite{fermions,ReferenceEnsembleFootnote}. 
It has also been shown that weak-localization effects (their clean
counterparts) can be related to corrections to scaling near a disordered 
(clean) Fermi liquid fixed point \cite{fermions}.
Let us generalize those considerations to include the effects of a small 
magnetic field. The scale dimension of
$\chi_t$ is zero\cite{fermions}, so in terms of a scaling function $F$
we have
\begin{mathletters}
\label{eqs:12}
\begin{equation}
\chi_t^{\rm (ref)}({\mathbf k},\ell_m^{-1},u) = 
             F(b{\mathbf k},\ell_m^{-1}b,ub^{[u]})\quad,
\label{eq:12a}
\end{equation}
with $\ell_m$ the magnetic length. The latter is determined perturbatively
as follows. A nonzero magnetization leads to a mass or frequency cutoff in
the soft modes that is given by a cyclotron frequency $\Omega_c$ with 
$m$ playing the role of the magnetic field. In clean (disordered) systems,
the wavenumber scales like $\Omega$ ($\Omega^{1/2}$). Scaling the
wavenumber with $\ell_m$ thus leads to $\ell_m\sim 1/m$ in clean systems, 
and $\ell_m\sim 1/\sqrt{m}$ in disordered ones.
$u$ represents the 
leading irrelevant variable near the fixed point. Its scale dimension,
$[u]$, is equal to $[u] = -(d-2)$ in disordered systems, and $[u] = -(d-1)$
in clean ones\cite{fermions,chi_s}. $b$ is
a RG length rescaling factor. In the paramagnetic phase, 
$\ell_m^{-1} = 0$, and Eq.\ (\ref{eq:12a}), with 
$b\sim \vert{\mathbf k}\vert^{-1}$, yields,
\begin{equation}
\chi_t^{\rm (ref)}({\mathbf k},0,u)\sim \chi_t 
              - c\vert{\mathbf k}\vert^{-[u]}\quad,
\label{eq:12b}
\end{equation}
with $c\sim u$. This is the nonanalyticity that leads to long-range 
interactions between the spin flucuations near the FM
phase transition. For $\ell_m^{-1}\neq 0$, $\chi_t^{\rm (ref)}$ is an 
analytic function of ${\mathbf k}^2$ and Eqs.\ (\ref{eq:12a},\ref{eq:12b}) give
\begin{equation}
\chi_t^{\rm (ref)}({\mathbf k},\ell_m^{-1},u)\sim \chi_t 
               - {c\,'}(m)\,{\mathbf k}^2\quad,
\label{eq:12c}
\end{equation}
with
\begin{eqnarray}
{c\,'}(m)\sim \ell_m^{2+[u]}\sim\cases{m^{-(2+[u])/2}\ \ &(disordered)\cr
                                   m^{-(2+[u])}\ \ &(clean)\cr}\ .
\label{eq:12d}
\end{eqnarray}
\end{mathletters}%
From this, with Eqs.\ (\ref{eqs:10}) and (\ref{eq:11}), we 
immediately obtain our main results, 
Eqs.\ (\ref{eqs:3}) (except for the nature of the leading
correction terms in Eq.\ (\ref{eq:1}), which we will discuss below). 
Note that for disordered systems the dimensionless parameter 
characterizing `small' wave numbers is $\vert{\mathbf k}\vert\ell\ll 1$, 
with $\ell$ the diffusive or transport mean-free path. The prefactors in 
Eqs.\ (\ref{eqs:3}) are hard to estimate, since they depend on
the value $\Gamma_t^{\rm ref}$ of $\Gamma_t$ in the fully renormalized 
reference ensemble\cite{ReferenceEnsembleFootnote}. For instance, for
the clean case in $d=3$ one finds, using the result of 
Ref.\ \onlinecite{chi_s}, 
${\tilde c}_3 = (32\pi/27)\,(N_{\rm F}\Gamma_t^{\rm ref})^2\,/\mu$.
Finally, we note that, at the level of the above scaling
argument, the analyticity of $\chi_t^{\rm (ref)}$
in powers of ${\mathbf k}^2$ for $\ell_m^{-1}\neq 0$ is an assumption.
We have checked this explicitly in perturbation theory,
verifying Eqs.\ (\ref{eq:12c},\ref{eq:12d}) using both a $Q$-matrix field 
theory\cite{fermions}, and standard many-body perturbation theory, and will
further discuss it from a RG point of view next.

We now show that the corrections to Eqs.\ (\ref{eqs:10}) that result
from taking into account the $\delta\rho\,$-fluctuations, as well as terms
of higher than Gaussian order in $\bpi$, cannot change the above results.
This is most easily done in the framework of the RG.
We assign scale dimensions $-1$ and $-z$ to lengths and
times, respectively, with $z$ the dynamical critical exponent, and scale
dimensions $[\pi_i(x)] = (d+z-2+\eta')/2$ and $[\rho(x)] = (d+z+\eta)/2$
to the fields. Then Eq.\ (\ref{eq:10a}) tells us that there is a
Gaussian fixed point with exponents
\begin{equation}
\eta = 2\quad,\quad \eta' = 0\quad,\quad z = 2\quad,
\label{eq:13}
\end{equation}
that describes a FM state. To check for
relevant operators that would destroy this fixed point, we
systematically expand Eq.\ (\ref{eq:8b}) in powers of
$\delta\rho$ and $\bpi$, and integrate out $\delta\rho$ perturbatively
to obtain an effective action in terms of $\bpi$. There are several
terms that dimensionally could lead to a $\vert{\mathbf k}\vert^{d-1}$ 
in the clean case and
a $\vert{\mathbf k}\vert^{d-2}$ in the disordered case in Eq.\ (\ref{eq:12c}), 
rather than a ${\mathbf k}^2$ with a coefficient
that is nonanalytic in $m$. In RG language, this would
be a relevant operator with respect to our Gaussian fixed point. However,
it turns out that there are Ward identities\cite{ZJ} that ensure, order
by order in the expansion in fluctuations of the order parameter, that
all terms of $O(\bpi^2)$, whether or not they couple to $\delta\rho$,
are multiplied by at least a gradient squared\cite{LongitudinalFootnote}.
We have also checked this by means of explicit perturbative calculations
for selected vertices. Similar arguments show that the second term on
the r.h.s. of Eq.\ (\ref{eq:11}) is the leading frequency dependence.
As a result, the Gaussian fixed point identified above is stable by
power counting. The leading nonanalytic correction to the
$\Omega \sim {\mathbf k}^2$ dispersion arises from renormalizations of the
Gaussian action due to terms of $O(\bpi^4)$. The resulting operators 
potentially have scale dimensions $2-d$ (disordered) and
$d-1$ (clean), respectively. This reflects the largest possible corrections
due to potentially soft modes; explicit calculations would be necessary
to ascertain whether or not terms of this order actually 
exist\cite{LongitudinalFootnote}.
We conclude that the Eqs.\ (\ref{eqs:12}) are asymptotically exact. 
The exact magnon dispersion relation is thus given by Eq.\ (\ref{eq:1}),
and the largest possible corrections are of 
$O(\vert{\mathbf k}\vert^{2+\zeta})$, with $\zeta$ from Eq.\ (\ref{eq:2}).

At $T>0$, temperature effects will compete with the
magnetization in protecting the weak-localization singularities, and their
clean counterparts, in the spin-triplet
channel\cite{R}. Therefore, for $m<T\ll T_{\rm F}$ in appropriate units, the
$m$ in the brackets in Eqs.\ (\ref{eqs:3}) will be replaced by $T$,
leading to a nonanalytic $T$-dependence of the coefficient in the
dispersion relation. Other consequences of a non-zero temperature
are more subtle because of the occurrence of multiple temperature 
scales\cite{us3}, and will be investigated separately in the future.

We conclude by discussing ways to experimentally verify our predictions.
To our knowledge, no systematic studies of the prefactor of the 
${\mathbf k}^2$-term in the dispersion relation have been performed.
Such a study should be easier to do for disordered
systems than for clean ones, since (1) the predicted effect is much
larger in the disordered case, and (2) in the disordered case it will
be easier to find a material near the FM quantum phase
transition (e.g. by fine tuning the concentration of the magnetic
ingredient in an alloy).

The most convincing experimental evidence would be an explicit measurement 
of the $m$-dependence of the dispersion relation. This would require measuring 
different samples with different values of the magnetization $m$, and 
extracting the $m$-dependence
from the measured inverse magnon masses $D(m)$. In a $3$-$d$
disordered system, $D(m)$ for small $m$ should scale like 
$m^{1/2}$ (instead of $m$ according to RPA-like theories). 
Another possibility is to measure a single sample with a small magnetization,
and to identify a {\em quantitative} difference of the measured 
magnon mass from that predicted by RPA-like theories. For instance, it
has been reported that in Fe and Ni that prefactor is larger than expected by a
factor of 2 to 3\cite{Moriya_p206}. Since the magnetization in these 
materials is not small, it is unlikely that this discrepancy is related
to the predicted effect. However, similar experiments on materials with
a small magnetization should suffice to corroborate or refute the present
theory.

This work was initiated at the Aspen Center for Physics, and supported by 
the NSF under grant numbers DMR--95--10185, DMR--96--32978, and DMR--97--07701 
and by the DFG under grant number Vo659/1--1.

\vfill\eject
\end{document}